\let\includefigures=\iftrue
%
%
%
\newfam\black
\input harvmac.tex
\input rotate
\input epsf
\input xyv2
\noblackbox
\includefigures
\message{If you do not have epsf.tex (to include figures),}
\message{change the option at the top of the tex file.}
\def\figin{\epsfcheck\figin}\def\figins{\epsfcheck\figins}
\def\epsfcheck{\ifx\epsfbox\UnDeFiNeD
\message{(NO epsf.tex, FIGURES WILL BE IGNORED)}
\gdef\figin##1{\vskip2in}\gdef\figins##1{\hskip.5in}
\else\message{(FIGURES WILL BE INCLUDED)}%
\gdef\figin##1{##1}\gdef\figins##1{##1}\fi}
\def\DefWarn#1{}
\def\N{{\cal N}}
\def\figinsert{\goodbreak\midinsert}
\def\ifig#1#2#3{\DefWarn#1\xdef#1{fig.~\the\figno}
\writedef{#1\leftbracket fig.\noexpand~\the\figno}%
\figinsert\figin{\centerline{#3}}\medskip\centerline{\vbox{\baselineskip12pt
\advance\hsize by -1truein\noindent\footnotefont{\bf
Fig.~\the\figno:} #2}}
\bigskip\endinsert\global\advance\figno by1}
\else
\def\ifig#1#2#3{\xdef#1{fig.~\the\figno}
\writedef{#1\leftbracket fig.\noexpand~\the\figno}%
\global\advance\figno by1} \fi
\def\yboxit#1#2{\vbox{\hrule height #1 \hbox{\vrule width #1
\vbox{#2}\vrule width #1 }\hrule height #1 }}
\def\fillbox#1{\hbox to #1{\vbox to #1{\vfil}\hfil}}
\def\ybox{{\lower 1.3pt \yboxit{0.4pt}{\fillbox{8pt}}\hskip-0.2pt}}

\def\rightarrowbox#1#2{
  \setbox1=\hbox{\kern#1{${ #2}$}\kern#1}
  \,\vbox{\offinterlineskip\hbox to\wd1{\hfil\copy1\hfil}
    \kern 3pt\hbox to\wd1{\rightarrowfill}}}

\def\half{{1\over 2}}
\def\Tr{{{\rm Tr~ }}}

\def\tilde{\widetilde}

\def\II{\relax{I\kern-.10em I}}

\def\IZ{\relax\ifmmode\mathchoice
{\hbox{\cmss Z\kern-.4em Z}}{\hbox{\cmss Z\kern-.4em Z}}
{\lower.9pt\hbox{\cmsss Z\kern-.4em Z}} {\lower1.2pt\hbox{\cmsss
Z\kern-.4em Z}}\else{\cmss Z\kern-.4em Z}\fi}
\def\IB{\relax{\rm I\kern-.18em B}}
\def\IC{{\relax\hbox{$\inbar\kern-.3em{\rm C}$}}}
\def\ID{\relax{\rm I\kern-.18em D}}
\def\IE{\relax{\rm I\kern-.18em E}}
\def\IF{\relax{\rm I\kern-.18em F}}
\def\IG{\relax\hbox{$\inbar\kern-.3em{\rm G}$}}
\def\IGa{\relax\hbox{${\rm I}\kern-.18em\Gamma$}}
\def\IH{\relax{\rm I\kern-.18em H}}
\def\II{\relax{\rm I\kern-.18em I}}
\def\IK{\relax{\rm I\kern-.18em K}}
\def\IN{\relax{\rm I\kern-.18em N}}
\def\IP{\relax{\rm I\kern-.18em P}}

\def\hat{\widehat}

\def\inbar{\,\vrule height1.5ex width.4pt depth0pt}

\font\cmss=cmss10 \font\cmsss=cmss10 at 7pt
\def\IR{\relax{\rm I\kern-.18em R}}

\def\id {{\bf 1}}

\def\tilde{\widetilde}

\def\lp10{l_P^{10}}
\def\lp11{l_P^{11}}
\def\R11{R_{11}}

\def\R{{\cal R}}

\def\tilde{\widetilde}

\def\ct{\tilde{c}}

%
%

\lref\KrausJV{ P.~Kraus, A.~V.~Ryzhov and M.~Shigemori, ``Loop
equations, matrix models, and N = 1 supersymmetric gauge
theories,'' arXiv:hep-th/0304138.
}

\lref\CachazoYC{ F.~Cachazo, N.~Seiberg and E.~Witten, ``Chiral
Rings and Phases of Supersymmetric Gauge Theories,'' JHEP {\bf
0304}, 018 (2003) [arXiv:hep-th/0303207].
}

\lref\CachazoZK{ F.~Cachazo, N.~Seiberg and E.~Witten, ``Phases of
N = 1 supersymmetric gauge theories and matrices,'' JHEP {\bf
0302}, 042 (2003) [arXiv:hep-th/0301006].
}

\lref\CachazoRY{ F.~Cachazo, M.~R.~Douglas, N.~Seiberg and
E.~Witten, ``Chiral rings and anomalies in supersymmetric gauge
theory,'' JHEP {\bf 0212}, 071 (2002) [arXiv:hep-th/0211170].
}

\lref\CachazoPR{ F.~Cachazo and C.~Vafa, ``N = 1 and N = 2
geometry from fluxes,'' arXiv:hep-th/0206017.
}

\lref\SeibergJQ{ N.~Seiberg, ``Adding fundamental matter to
'Chiral rings and anomalies in  supersymmetric gauge theory',''
JHEP {\bf 0301}, 061 (2003) [arXiv:hep-th/0212225].
}

\lref\CachazoJY{ F.~Cachazo, K.~A.~Intriligator and C.~Vafa,
Nucl.\ Phys.\ B {\bf 603}, 3 (2001) [arXiv:hep-th/0103067].
}

\lref\IntriligatorJR{ K.~A.~Intriligator, R.~G.~Leigh and
N.~Seiberg, ``Exact superpotentials in four-dimensions,'' Phys.\
Rev.\ D {\bf 50}, 1092 (1994) [arXiv:hep-th/9403198].
}

\lref\AldayGB{ L.~F.~Alday and M.~Cirafici, ``Effective
superpotentials via Konishi anomaly,'' arXiv:hep-th/0304119.
}

\lref\FerrariYR{ F.~Ferrari, ``Quantum parameter space in super
Yang-Mills. II,'' Phys.\ Lett.\ B {\bf 557}, 290 (2003)
[arXiv:hep-th/0301157].
}

\lref\WittenFP{ E.~Witten, ``An SU(2) Anomaly,'' Phys.\ Lett.\ B
{\bf 117}, 324 (1982).
}

\lref\ChoBI{ P.~L.~Cho and P.~Kraus, ``Symplectic SUSY gauge
theories with antisymmetric matter,'' Phys.\ Rev.\ D {\bf 54},
7640 (1996) [arXiv:hep-th/9607200].
}

\lref\CsakiEU{ C.~Csaki, W.~Skiba and M.~Schmaltz, ``Exact results
and duality for Sp(2N) SUSY gauge theories with an  antisymmetric
tensor,'' Nucl.\ Phys.\ B {\bf 487}, 128 (1997)
[arXiv:hep-th/9607210].
}

\lref\KrausJF{ P.~Kraus and M.~Shigemori, ``On the matter of the
Dijkgraaf-Vafa conjecture,'' JHEP {\bf 0304}, 052 (2003)
[arXiv:hep-th/0303104].
}

\lref\WittenYE{ E.~Witten, ``Chiral ring of Sp(N) and SO(N)
supersymmetric gauge theory in four  dimensions,''
arXiv:hep-th/0302194.
}

\lref\ElitzurGK{ S.~Elitzur, A.~Forge, A.~Giveon,
K.~A.~Intriligator and E.~Rabinovici, ``Massless Monopoles Via
Confining Phase Superpotentials,'' Phys.\ Lett.\ B {\bf 379}, 121
(1996) [arXiv:hep-th/9603051].
}

\lref\AganagicXQ{ M.~Aganagic, K.~Intriligator, C.~Vafa and
N.~P.~Warner, ``The glueball superpotential,''
arXiv:hep-th/0304271.
}

\lref\VenezianoAH{ G.~Veneziano and S.~Yankielowicz, ``An
Effective Lagrangian For The Pure N=1 Supersymmetric Yang-Mills
Theory,'' Phys.\ Lett.\ B {\bf 113}, 231 (1982).
}

\lref\IntriligatorAU{ K.~A.~Intriligator and N.~Seiberg,
``Lectures on supersymmetric gauge theories and electric-magnetic
duality,'' Nucl.\ Phys.\ Proc.\ Suppl.\  {\bf 45BC}, 1 (1996)
[arXiv:hep-th/9509066].
}

\lref\FerrariKQ{ F.~Ferrari, ``Quantum parameter space and double
scaling limits in N = 1 super  Yang-Mills theory,'' Phys.\ Rev.\ D
{\bf 67}, 085013 (2003) [arXiv:hep-th/0211069].
}

\lref\DijkgraafDH{ R.~Dijkgraaf and C.~Vafa, ``A perturbative
window into non-perturbative physics,'' arXiv:hep-th/0208048.
}

\lref\DijkgraafVW{ R.~Dijkgraaf and C.~Vafa, ``On geometry and
matrix models,'' Nucl.\ Phys.\ B {\bf 644}, 21 (2002)
[arXiv:hep-th/0207106].
}

\lref\DijkgraafFC{ R.~Dijkgraaf and C.~Vafa, ``Matrix models,
topological strings, and supersymmetric gauge theories,'' Nucl.\
Phys.\ B {\bf 644}, 3 (2002) [arXiv:hep-th/0206255].
}

\lref\TerningTH{ J.~Terning, ``TASI-2002 lectures:
Non-perturbative supersymmetry,'' [arXiv:hep-th/0306119].
}

\lref\KonishiTU{ K.~i.~Konishi and K.~i.~Shizuya, ``Functional
Integral Approach To Chiral Anomalies In Supersymmetric Gauge
Theories,'' Nuovo Cim.\ A {\bf 90}, 111 (1985).
}

\lref\KonishiHF{ K.~Konishi, ``Anomalous Supersymmetry
Transformation Of Some Composite Operators In Sqcd,'' Phys.\
Lett.\ B {\bf 135}, 439 (1984).
}

\lref\DijkgraafXD{ R.~Dijkgraaf, M.~T.~Grisaru, C.~S.~Lam, C.~Vafa
and D.~Zanon, ``Perturbative computation of glueball
superpotentials,'' arXiv:hep-th/0211017.
}

\newbox\tmpbox\setbox\tmpbox\hbox{\abstractfont }
\Title{\vbox{\baselineskip12pt\hbox to\wd\tmpbox{\hss
hep-th/0307063}}} {\vbox{\centerline{Notes on supersymmetric
$Sp(N)$ theories with an}
\smallskip
\centerline{antisymmetric tensor}}}
\smallskip
\centerline{ Freddy Cachazo}
\smallskip
\bigskip
\centerline{School of Natural Sciences, Institute for Advanced
Study, Princeton NJ 08540 USA}
\medskip
\bigskip
\vskip 1cm
 \noindent

We study supersymmetric $Sp(N)$ gauge theories with an
antisymmetric tensor and degree $n+1$ tree level superpotential.
The generalized Konishi anomaly equations derived in
hep-th/0304119 and hep-th/0304138 are used to compute the low
energy superpotential of the theory. This is done by imposing a
certain integrality condition on the periods of a meromorphic one
form. Explicit computations for $Sp(2)$, $Sp(4)$, $Sp(6)$ and
$Sp(8)$ with cubic superpotential are done and full agreement with
the results of the dynamically generated superpotential approach
is found. As a byproduct, we find a very precise map from $Sp(N)$
to a $U(N+2n)$ theory with one adjoint and a degree $n+1$ tree
level superpotential.

\Date{July, 2003}

\newsec{Introduction}

$\N=1$ supersymmetric gauge theories have provided a very rich
arena to explore nonperturbative quantum effects. Holomorphy has
been the key tool in controlling the strong coupling dynamics. In
the 90's, these techniques were extensively used to determine the
structure of holomorphic quantities in the infrared of
asymptotically free theories (for a review, see e.g.
\refs{\IntriligatorAU,\TerningTH}). For pure $\N = 1$ $SU(N)$ a
massive vacuum is generated with low energy superpotential
\eqn\loon{ W_{\rm low} = N (\Lambda^{3N})^{1/N}.}
This has the same low energy physics as the proposal of Veneziano
and Yankielowicz \VenezianoAH. They introduced a single scalar
superfield $S$ which is a fundamental field in the infrared but it
is composite in the microscopic theory. In the latter, $S$ is
given by the glueball superfield $-{1\over 32\pi^2}\Tr
W_{\alpha}W^{\alpha}$. In the infrared, the theory develops an
effective superpotential,
\eqn\vysu{ W_{\rm VY} = S\left[ \ln \left( {\Lambda^{3N}\over
S^N}\right) + N \right].}
At the extremum of \vysu, the fundamental field $S$ acquires an
expectation value $S^N = \Lambda^{3N}$ and \vysu\ reduces to
\loon. Clearly, the only contribution to this vev is
nonpertubative and in perturbation theory we can write $S^N_{\rm
p.t.} = 0$. It was proven in \CachazoRY\ that in the microscopic
theory, $S$, defined as the glueball superfield, satisfies the
identity $S^N \simeq 0$, where $\simeq$ means valid in the chiral
ring of the theory. In \CachazoRY, it was also suggested that the
exchange of the identity $S^N\simeq 0$ in the microscopic theory
by the field equation $S^N_{\rm p.t.} = 0$ in the effective theory
is a sign of dual descriptions. If this is the case, no
information of the constraint $S^N_{\rm p.t} = 0$ should be
visible in the off-shell low energy effective superpotential for
$S$.

In general, for any pure $\N =1$ gauge theory, with a simple Lie
group $G$, the Veneziano-Yankielowicz superpotential is given by
\eqn\gvy{ W_{\rm VY} = S\left[ \ln \left( {\Lambda^{3h}\over
S^h}\right) + h \right] }
where $h$ is the dual Coxeter number of the Lie algebra of $G$.

In \CachazoRY, it was conjectured that the effective equation of
motion $S^h = \Lambda^{3h}$ or $S^h_{\rm p.t.} = 0$ was reproduced
as an identity in the chiral ring of the microscopic theory, i.e.
$S^h \simeq 0$. This was proven for $SO(N)$ and $Sp(N)$ gauge
groups in \WittenYE.

Recently, a new way of computing low energy holomorphic physics of
a large class of $\N=1$ gauge theories using bosonic matrix models
was conjectured \refs{\DijkgraafFC,\DijkgraafVW,\DijkgraafDH}.
This conjecture motivated the development of superfield techniques
\DijkgraafXD\ and of generalized Konishi anomalies \CachazoRY\ to
prove it.

Using the superfield techniques \KrausJF, and using the
generalized Konishi anomalies \refs{\KrausJV,\AldayGB}, the low
energy superpotential for $Sp(N)$ gauge theories with one matter
field in the antisymmetric representation and tree level
superpotential was computed. These theories were also studied in
the 90's using holomorphy to determine the dynamically generated
superpotentials \refs{\ChoBI,\CsakiEU}. In the far IR, after all
fields have been integrated out, the two approaches should give
the same answer. By this we mean a single function $W_{\rm low}$,
the low energy superpotential, that depends on the dynamically
generated scale $\Lambda$ and tree level superpotential
parameters.

However, a discrepancy was found in \KrausJF\ for several
examples, namely, $Sp(4)$, $Sp(6)$ and $Sp(8)$ in the unbroken
classical vacuum with a cubic superpotential. The difference
always sets in at order $\Lambda^{3h}$ where $h=N/2+1$ is the dual
Coxeter number of $Sp(N)$. It was then suggested that perhaps $S$
satisfies relations coming from its glueball origin. This
explanation, however, would contradict the dual picture mentioned
above.

In \AganagicXQ, a possible explanation to the discrepancy was
suggested. An ambiguity in the UV behavior of a theory was
discovered if one allows for supergroups. A prescription to
``F-term complete" a theory was given by thinking about the gauge
group $G(N)$ as embedded in a supergroup $G(N+k|k)$ with $k\to
\infty$. Moreover, the matrix model, glueball superpotential, or
Konishi anomaly computations, were claimed to be computing the
superpotential of the F-completion $G(N+k|k)$. Therefore, the
analysis of \refs{\KrausJF,\KrausJV}\ and \AldayGB, which was
compared to a standard $Sp(N)$ field theory calculation
\refs{\ChoBI,\CsakiEU}, had actually being done for $Sp(N)\times
Sp(0)$. Here $Sp(0)$ is a remnant of the F-completion
$Sp(N+k_1+k_2|k_1+k_2)$ broken down to $Sp(N+k_1| k_1)\times
Sp(k_2|k_2)$. This completion produces residual instantons that
are not present in the standard UV completion of $Sp(N)$.

In these notes we show that by using the generalized Konishi
anomaly equations derived in \KrausJV\ and \AldayGB\ and imposing
that the periods of the generating function of chiral operators,
$T(z)=\langle \Tr {1\over z-\Phi}\rangle$, satisfy
\eqn\cond{{1\over 2\pi i}\oint_{A_1} T(z)dz = N, \qquad  {1\over
2\pi i}\oint_{A_2} T(z)dz = 0,}
we get full agreement with the dynamically generated
superpotential approach \refs{\ChoBI,\CsakiEU}, i.e, no
discrepancy was observed at any order computed in the examples. In
particular, these contain the examples considered in \KrausJF.

The $A_k$'s in \cond\ refer to the $A$-cycles of a genus $g=1$
Riemann surface
\eqn\rrss{ y^2 = W'(z)^2 + f(z)}
where $W'(z) = 0 $ is the classical F-term field equation.

The IR dynamics is taken into account by imposing that the period
of $T(z)dz$ over the $B$-cycle of \rrss\ be an integer
\refs{\CachazoPR,\CachazoYC}.

We find that the cut $A_2$ {\it does not} close up on-shell. This
is very surprising at first due to the second equation in \cond.
The fact that the vanishing of the period of $T(z)dz$ through a
given cut implies that the cut disappears on-shell has been
usually assumed for $U(N)$ theories with excellent results so far
(see e.g. \refs{\FerrariKQ,\CachazoRY}). Here we show that this
assumption is not valid in general and it is the reason for the
discrepancy observed in \refs{\KrausJF,\KrausJV,\AldayGB}.

The main difference between $Sp(N)$ with antisymmetric tensor and
$U(N)$ with adjoint is that in the former the branch points of
\rrss\ are simple poles of $T(z)dz$ while in the latter they are
regular points.

We also consider general degree $n+1$ superpotentials and generic
classical vacua. In this case, the IR dynamics is also taken into
account by imposing that the periods over the $B$-cycles of \rrss,
which in general has genus $n-1$, be integers
\refs{\CachazoPR,\CachazoYC}.

The actual computation is carried out by mapping the problem to a
$U(N+2n)$ theory. This is a very precise map which might suggest a
new kind of duality.

To give an example, $Sp(N)$ with one antisymmetric tensor\foot{We
follow the convention used in \KrausJF\ where $N$ is an even
integer, and $Sp(2)\simeq SU(2)$.} corresponds to $U(N+2n)$ with
one adjoint. If $Sp(N)$ is classically unbroken, then $U(N+2n)$ is
classically broken to $U(N+2)\times U(2)^{n-1}$. Quantum
mechanically, the vacua of $Sp(N)$ correspond to $U(N+2n)$ vacua
with confinement index $t=2$ (defined in \CachazoZK ). Given that
$t=2$, the computation is effectively carried out in $U(N/2+n)$
classically broken to $U(N/2+1)\times U(1)^{n-1}$. In particular,
for a cubic superpotential $n=2$ and $U(N/2+2)$ is broken to
$U(N/2+1)\times U(1)$. The $U(1)$ factor enters in the low energy
superpotential starting at order $\Lambda^{3(N/2+1)}$. This
explains why the discrepancy found in \KrausJF\ started at
different orders for different values of $N$.

The presence of the $U(1)$ factors is reminiscent of the $Sp(0)$
factors in \AganagicXQ. However, it is important to note that
these $U(1)$ factors are forced upon us and are crucial to the
agreement with the standard $Sp(N)$ dynamically generated
superpotential results of \refs{\ChoBI,\CsakiEU}.

This paper is organized as follows: In section 2 we consider
$Sp(N)$ theories with one antisymmetric tensor in detail. The
anomaly equations are solved for a general degree $n+1$
superpotential. The mapping to $U(N+2n)$ is shown and used to
compute $W_{\rm low}$. In section 3, the low energy superpotential
for $Sp(4)$, $Sp(6)$ and $Sp(8)$ is computed. Also, as a
consistency check, $Sp(0)$ and $Sp(2)$ are considered. In section
4, we end with some conclusions and open questions. In appendix A,
we review and compute relevant $U(N+2n)$ results. Finally, in
appendix B, we review the dynamically generated superpotential
computation of $W_{\rm low}$ for $Sp(4)$, $Sp(6)$ and $Sp(8)$.

\newsec{$Sp(N)$ with antisymmetric matter}

We consider $\N=1$ supersymmetric $Sp(N)$ gauge theories with a
chiral superfield in the antisymmetric representation $\Phi \equiv
A J$. $A$ is a $N\times N$ antisymmetric matrix and $J$ is the
invariant antisymmetric tensor of $Sp(N)$ , i.e., $\id_{N/2\times
N/2} \times i\sigma_2$.

The theory is deformed by a degree $n+1$ tree level superpotential
\eqn\sup{ W_{\rm tree} = \sum_{k=0}^{n}{g_k\over k+1}\Tr
\Phi^{k+1}.}
Note that we have explicitly included the term $g_0 \Tr \Phi$. In
the traceless case, $g_0$ can be used as a Lagrange multiplier
imposing the tracelessness constraint, $\langle \Tr \Phi \rangle =
0$.

Classically, the gauge group is generically broken to
$Sp(N_1)\times \ldots \times Sp(N_n)$ with $N= \sum_{i=1}^n N_i$
and $N_i$ even.

We are interested in the generating functions of chiral operators
\refs{\CachazoRY,\KrausJV,\AldayGB},
\eqn\chi{ T(z) = \Tr {1\over z-\Phi}, \quad R(z)=-{1\over 32
\pi^2}\Tr {W_{\alpha}W^{\alpha}\over z-\Phi}. }

Following \CachazoRY, a set of equations which constraint the
generating functions \chi\ was derived in
\refs{\KrausJV,\AldayGB}. These equations are obtained by studying
a generalization of the Konishi anomaly
\refs{\KonishiHF,\KonishiTU} and they can be used to determine
$T(z)$ and $R(z)$ up to $2n$ complex parameters.

\subsec{Anomaly equations and solution}

The equations are given by \refs{\KrausJV,\AldayGB},
\eqn\ano{\eqalign{ [W'(z)R(z)]_{-} =\; & \half R^2(z), \cr
              [W'(z)T(z)]_{-} =\; & T(z)R(z)+2{d\over dz}R(z). }}

The first equation gives $R(z)$ in terms of $W'(z)= \sum_{k=0}^n
g_k z^k$ and a degree $n-1$ polynomial $f(z)\equiv -
2[W'(z)R(z)]_+$ as follows,
\eqn\soone{R(z) = W'(z) -\sqrt{W'(z)^2+f(z)}.}
The branch of the square root is chosen so that $R(z)\sim 1/z$ for
$z\to \infty$. This is the correct classical behavior since only
the first sheet is visible classically and the appearance of a
second sheet is a quantum effect.

Classically, $R(z)$ is a meromorphic function on a sphere. Quantum
mechanically, it is a meromorphic function on a Riemann surface of
genus $n-1$ defined by
\eqn\riesf{ y^2 = W'(z)^2 + f(z).}
Let us denote by $A_i$ and $B_i$, with $i=1,\ldots ,n-1$,
canonical basis of 1-cycles in \riesf. $A_i$ is defined to circle
the i$^{\rm th}$ branch cut. The $A_n$ cycle can also be defined
as the cycle around the n$^{\rm th}$ cut. This cycle is not
relevant when only holomorphic differentials are considered but
here we will be dealing with meromorphic differentials.

The equation for $T(z)$ can be solved in terms of $W'(z)$, $f(z)$,
and a degree $n-1$ polynomial $c(z)\equiv [W'(z)T(z)]_+$,
\eqn\sotw{T(z) = {c(z)\over y(z)} + 2 {W''(z)\over y(z)} - 2
{y'(z)\over y(z)}.}
Note that the combination $\ct (z) = c(z)+ 2 W''(z)$ is again a
polynomial of degree $n-1$. In addition, note that the last term
in \sotw\ can be written as a logarithmic derivative. Combining
these observations \sotw\ becomes
\eqn\obw{ T(z)  = {\ct(z)\over y(z)} - {d\over dz}\log \left(
W'(z)^2 + f(z) \right).}
{}From \obw\ we learn that $T(z)dz$ is a meromorphic differential
on the Riemann surface \riesf. It has simple poles at $z=\infty$
on the first and second sheets, and in all branch points. $T(z)dz$
depends on $2n$ complex parameters given by the coefficients of
$f(z)$ and $\ct(z)$.

The $n$ coefficients in $\ct(z)$ can be fixed by imposing
\eqn\peru{ {1\over 2\pi i}\oint_{A_i}T(z)dz = N_i \quad {\rm for}
\quad i=1,\ldots ,n.}
The n$^{th}$ $A$-cycle has to be included since $T(z)dz$ has a
pole at infinity. In the unbroken classical vacuum $N_1 =N$ and
$N_i = 0$ for $i=2,\ldots ,n$ .

On the other hand, $n-1$ of the coefficients in $f(z)$ can be
expressed in terms of
\eqn\bpe{  {1\over 2\pi i}\oint_{B_i} T(z)dz = b_i \quad {\rm for}
\quad i=1, \ldots , n-1.}
The n$^{th}$ parameter in $f(z)$ is related to the scale of the
theory, $\Lambda$, and depends on the regularization scheme.

The parameters $b_i$'s are determined by the IR dynamics. In
\CachazoYC, it was shown that for a $U(\tilde N)$ theory with
adjoint and fundamental matter, solving the field equations of the
low energy effective superpotential is equivalent to imposing the
integrality of the periods $b_i$'s in \bpe.

Here we assume that the IR dynamics of the $Sp(N)$ theory is also
determined by imposing the integrality of $b_i$. This surprisingly
simple condition seems to be a generic feature of theories where a
hyperelliptic Riemann surface emerges. It would be interesting to
explore this in more detail. We comment more on this issue in
section 4.

At this point we have all the ingredients to compute $T(z)$ and
$W_{\rm low}$. However, it proves more convenient to map this
problem to a known one. As a byproduct we find a surprisingly
precise relation to a $U(N+2n)$ theory.

\subsec{ Mapping of $Sp(N)$ to $U(N+2n)$}

Consider $U(\tilde N)$ with a single adjoint chiral superfield
$\Phi_U$ and tree level superpotential
\eqn\duasu{ W_{\rm tree(U)} = \sum_{k=0}^{n}{h_k\over k+1}\Tr
\Phi_U^{k+1}.}
The generalized Konishi anomaly equations for this theory are
given by \CachazoRY,
\eqn\unko{[W'_U(z)R_U(z)]_- = R^2_U(z), \qquad [W'_U(z)T_U(z)]_- =
2R_U(z) T_U(z). }
The generating functions of chiral operators
\eqn\erlw{ R_U(z) = -{1\over 32\pi^2 }\Tr
{W_{U\alpha}W_U^{\alpha}\over z-\Phi_U}, \qquad T_U(z) = \Tr {1
\over z-\Phi_U }}
are given in terms of two degree $n-1$ polynomials $f_U(z) =
-4[W'_U(z)R_U(z)]_+$ and $c_U(z) = [W'_U(z)T(z)]_+$ as follows
\eqn\ppqq{ T_U(z)  = {c_U(z)\over y_U(z)}, \qquad R_U(z) = \half
\left( W'_U(z) - y_U(z)\right).}
These are meromorphic functions on the Riemann surface $y_U^2 =
W'_U(z)^2 + f_U(z)$.

As mentioned before, the IR dynamics of this theory is determined
by simply imposing the integrality of all the periods of $T_U(z)
dz$.

Motivated by the form of the equations \ano\ and \unko\ we propose
the following map,
\eqn\mappro{W'_U(z) = 2 W'(z), \qquad R_U(z) = R(z) . }
{}From these equations it is easy to get that
\eqn\easyy{ f_U(z) = 4 f(z), \qquad y_U(z) = 2y(z).}

The key observation is that $T(z)$ for $Sp(N)$ given by \obw\ has
the same form as $T_U(z)$ in \ppqq\ except for a logarithmic
derivative of a meromorphic function on \riesf. Therefore, the
extra term does not affect the dynamics since it has integer
periods automatically.

This leads us to identify
\eqn\tids{T_U(z) = T(z) + {d\over dz}\log\left( W'(z)^2 + f(z)
\right) }
or equivalently $c_U(z) = 2 \ct (z)$. It is not difficult to check
that these identifications are consistent as we will see next.

\bigskip
\noindent{\it Chiral operators}

It is important to get the relation between the $Sp(N)$ invariants
$\Tr \Phi^k$ with $k\geq 0$ and the $U(\tilde N)$ invariants $\Tr
\Phi_U^k$ with $k\geq 0$. This is done by solving \tids\ order by
order in a $1/z$ expansion. Let us rewrite \tids\ more explicitly,
solving for $T(z)$,
\eqn\consk{\left\langle \Tr {1\over z-\Phi} \right\rangle =
\left\langle \Tr {1\over z-\Phi_U} \right\rangle - {d\over
dz}\log\left( W'(z)^2 + f(z) \right) .}
Note that the first $n$ orders in \consk, i.e., up to $1/z^{n}$
are equivalent to
\eqn\secck { [W'(z)T(z)]_+ + 2W''(z) =  [W'(z)T_U(z)]_+.}
Therefore, we do not get either extra information or an
inconsistency by studying $c_U(z) = 2 \ct (z)$.

Of particular interest is the leading order in \consk,
\eqn\lead{ \Tr \id = \Tr \id_U - 2n. }
This gives $\tilde N = N + 2n$ as claimed in the introduction.

Let us also mention a relation which follows from \mappro\ and
will play a key role in the sequel
\eqn\oruf{ S_U = S.}

\bigskip
\noindent{\it Classical breaking pattern}

In order to complete the map we have to identify the correct
classical breaking pattern. A $U(\tilde N)$ theory with adjoint
$\Phi_U$ and tree level superpotential \duasu\ generically has
classical vacua with unbroken $U(\tilde N_1)\times \ldots \times
U(\tilde N_n)$. The relation between $\tilde N_i$ and $N_i$ can be
found by studying the periods of $T_U(z)$ through the $A$-cycles.

In the classical vacuum where $Sp(N)$ is broken to $Sp(N_1)\times
\ldots \times Sp(N_n)$ we have \peru,
\eqn\perut{ {1\over 2\pi i}\oint_{A_i}T(z)dz = N_i \quad {\rm for}
\quad i=1,\ldots ,n.}
On the other hand, the logarithmic derivative $\psi (z) ={d\over
dz}\log \left( W'(z)^2 + f(z) \right)$ satisfies
\eqn\loddw{ {1\over 2\pi i}\oint_{A_i}\psi(z)dz = 2 \quad {\rm
for} \quad i=1,\ldots ,n. }
Therefore,
\eqn\perufe{ {\tilde N}_i = {1\over 2\pi i}\oint_{A_i}T_U(z) dz =
{1\over 2\pi i}\oint_{A_i}T(z) dz + {1\over 2\pi
i}\oint_{A_i}\psi(z) dz = N_i + 2 \quad {\rm for} \quad i=1,\ldots
,n.}
Note that this is consistent with $\tilde N = \sum_{i=1}^n \tilde
N_i$.

In the $Sp(N)$ unbroken vacuum, $U(N+2n)$ is broken to
$U(N+2)\times U(2)^{n-1}$.

\bigskip
\noindent{\it Vacua identification}

The last step in defining the associated $U(\tilde N)$ problem is
the identification of vacua in the quantum theory. Recall that in
our normalization of $Sp(N)$, $N$ is an even number. Therefore,
the set of numbers $\tilde N_i = N_i + 2$ with $i=1,\ldots ,n$ has
$2$ as common divisor. Equivalently, if $m$ is the maximum common
divisor of the set $\tilde N_i$, $i=1,\ldots ,n$, then $m$ is
even.

The number of vacua for $U(N+2n)$ broken to $\prod_{i=1}^n
U(N_i+2)$ is $\prod_{i=1}^n (N_i+2)$. As shown in \CachazoZK,
these vacua fall into separate phases distinguished by the
``confinement index" $t$ ( defined in \CachazoZK ). $t$ takes
values in the set of integer factors of $m$. As mentioned above,
$m$ is always even. Therefore, at least two possibilities are
present in this theory: $t=1$ and $t=2$.

A very useful fact about vacua with confinement index $t> 1$ is
the following. $T_U(z)$ evaluated at vacua of $U(N+2n)$ with
classical breaking $U(N_1+2)\times \ldots \times U(N_n+2)$ and
confinement index $t$ can be effectively computed in a $U((N +
2n)/t)$ theory with classical breaking $U((N_1+2)/t)\times \ldots
\times U((N_n+2)/t)$ and same tree level superpotential. Let us
denote by $T_u(z)$ the generating function of chiral operators of
such a $U((N+2n)/t)$ theory. The relation between the two is very
simple \CachazoZK,
\eqn\hjps{ T_U(z) = t \; T_u(z).}

Consider for example the unbroken $Sp(N)$ vacuum. This theory is
maximally confining. Clearly $m=2$ since $N_i=0$ for $i=2,\ldots
,n$. Note also that $U(N+2n)$ vacua with $t=1$ are in a non
confining phase. Therefore $t$ must be equal to $2$. Motivated by
this result, we propose that in general $t=2$. This is consistent
with the fact that the center of $Sp(N)$ is $\IZ_2$.

Let us check this proposal by counting the number of vacua in the
general case. Pure supersymmetric $Sp(N_i)$ is expected to have
$h_i=N_i/2+1$ vacua distinguished by the solutions to the
effective superpotential \gvy, i.e., $S_i^{h_i} =
\Lambda_i^{3h_i}$. Therefore, when $Sp(N)$ is classically broken
to $\prod_{i=1}^n Sp(N_i)$ we find $\prod_{i=1}^n h_i$ vacua. In
the $U(N+2n)$ theory we have $\prod_{i=1}^n (N_i+2)$ which is
$2^n$ times larger. However, the number of vacua with $t=2$ is
less and it is computed in $U(N/2+n)$ broken to
$\prod_{i=1}^nU(N_i/2+1)$. Therefore, the number of vacua with
$t=2$ is $\prod_{i=1}^{n}(N_i/2+1)$ which is the desired answer.

However, this counting is not completely correct. The reason is
that for each vacuum of $U(N/2+n)$ there are $t=2$ vacua in
$U(N+2n)$ \CachazoZK. This would lead to twice the expected
number. The way this happens is through the relation between the
scales of the theories \refs{\CachazoJY,\CachazoZK}. If we denote
by $\Lambda_u$ and $\Lambda_U$ the scales of $U(N/2+n)$ and
$U(N+2n)$ respectively, then
\eqn\scalqw{ \Lambda^{N+2n}_u = \eta \Lambda^{N+2n}_U \quad {\rm
with} \quad \eta^t=1.}

As we will see next, in the case of a cubic superpotential and
unbroken $Sp(N)$ classical vacuum this puzzle is resolved because
the $Sp(N)$ low energy superpotential only depends on
$\Lambda_u^{2(N+2n)}$ and therefore it is invariant under the
$\IZ_2$ action $\eta\to -\eta$. We believe that this is also true
in the general case. We leave the proof of this for future work.

\subsec{Low energy superpotential}

The low energy superpotential, $W_{\rm low}$, is a single function
of the dynamically generated scale of the theory $\Lambda$ and the
tree level superpotential parameters. $W_{\rm low}$ is defined so
that the expectation value of the chiral operators $\Tr \Phi^k$
and $S =-{1\over 32\pi^2}\Tr W_{\alpha}W^{\alpha}$ are given by
\IntriligatorJR,
\eqn\kqwp{\eqalign{ {\del W_{\rm low}\over \del g_k} =\ & {1\over
k+1} \langle \Tr \Phi^{k+1} \rangle, \cr {\del W_{\rm low}\over
\del \log \Lambda_H}  =\ & (N+4) S.}}
The factor $(N+4)$ in the last equation is the coefficient of the
holomorphic beta function of $Sp(N)$ with $\Phi$ an antisymmetric
tensor. We denote by $\Lambda_H$ the scale of the high energy
theory before $\Phi$ is integrated out.

The equations \kqwp\ can be used to find $W_{\rm low}$ up to an
irrelevant constant independent of the couplings in the
superpotential and $\Lambda_H$.

For simplicity, and also because it is the case used in the
examples, let us consider a cubic superpotential
\eqn\cusp{ W_{\rm tree} = {g\over 3}\Tr \Phi^3 + {m\over 2}\Tr
\Phi^2 + \lambda \Tr\Phi.}
Using \consk\ we get
\eqn\cubr{ \eqalign{ \langle \Tr \id \rangle = \langle \Tr \id_U
\rangle - 4 , \quad & \langle \Tr \Phi \rangle = \langle \Tr
\Phi_{U} \rangle + 2 {m\over g} , \cr \langle \Tr \Phi^2 \rangle =
\langle \Tr \Phi^2_{U} \rangle -2\left( {m^2\over g^2}
-2{\lambda\over g} \right) , \quad & \langle \Tr \Phi^3 \rangle =
\langle \Tr \Phi^3_{U} \rangle -2\left( 3{S\over g} - {m^3\over
g^3} + 3 {m\lambda\over g^2} \right) .} }

These equations, together with \kqwp, determine $W_{\rm low}$ once
the corresponding vev's of $U(N+4)$ are known.

An important consistency check is the integrability of \kqwp. We
assume that the $U(N+4)$ problem has been solved, i.e., $W_{\rm
low (U)}$ is known. More explicitly, we assume that the system
\eqn\sukp{\eqalign{ {\del W_{\rm low(U)}\over \del h_k} =\ &
{1\over k+1} \langle \Tr \Phi_U^{k+1} \rangle, \cr {\del W_{\rm
low(U)}\over \del \log \Lambda_U}  =\ & 2(N+4) S_U } }
has been integrated. Recall that $h_2 = 2g$, $h_1 = 2m$, and
$h_0=2\lambda$ are the couplings of $W_{\rm tree(U)}$ \duasu.

Let us propose an ansatz for $W_{\rm low}$,
\eqn\ansz{W_{\rm low}(\Lambda_H , g ,m,\lambda )= {1\over 2}W_{\rm
low (U)}(\Lambda_U(\Lambda_H , g), 2g , 2m , 2\lambda ) -{1\over
3}{m^3\over g^2} + 2{\lambda m\over g}.}
The ansatz for $\Lambda_U = \Lambda_U(g,\Lambda_H)$ implies, by
dimensional analysis, that $\Lambda_U$ is proportional to
$\Lambda_H$ up to a $g$ dependent factor. This can be shown by
taking a derivative with respect to $\log \Lambda_H$ of \ansz.
This leads to
\eqn\logH{ {\del W_{\rm low}\over \del \log\Lambda_H} = (N+4)S_U
{\Lambda_H\over \Lambda_U}{ \del \Lambda_U \over \del \Lambda_H}.}
Using the last equation in \kqwp\ together with \oruf\ i.e.,
$S=S_U$, we get that the equation is satisfied by the ansatz.

It is straightforward to check that the equations for $\langle \Tr
\Phi \rangle$ and $\langle \Tr \Phi^2 \rangle$ in \cubr\ are
satisfied. The equation for $\langle \Tr \Phi^3 \rangle$ is more
interesting and leads to the following equation,
\eqn\morin{ {\del W_{\rm low(U)}\over \del \Lambda_U}{\del
\Lambda_U\over \del g} = - 4 {S\over g}.}
Using \oruf\ and the second equation in \sukp\ we get,
\eqn\ffnn{ \Lambda_U (\Lambda_H , g) =  g^{-{2\over
N+4}}\Lambda_H. }
up to a numerical constant. This $g$ dependence will play a
crucial role in the next section. In the unbroken $Sp(N)$ vacuum
the leading order of $W_{\rm low}$ does not depend on $g$. The
reason is that no $W$-boson is massive and has to be integrated
out in this vacuum. On the other hand, in $U(N+4)$ broken to
$U(N+2)\times U(2)$, massive $W$-bosons are integrated out. It is
very satisfying that the power of $g$ in \ffnn\ is precisely
correct to cancel the $g$ dependence in the $U(N+4)$ answer as
shown in the next section.

For future reference let us write the threshold matching relation
between the scales $\Lambda_H$ and $\Lambda$, the pure $Sp(N)$
scale after integrating out the antisymmetric tensor,
\eqn\mapu{ \Lambda^{3(N+2)} = \Lambda_H^{2(N+4)} m^{N-2}. }

Recall that $W_{\rm low (U)}$ in \ansz\ has to be computed around
the classical vacuum where $U(N+4)$ is broken to $U(N+2)\times
U(2)$ and evaluated in the quantum vacua with confinement index
$t=2$. This means that $W_{\rm low (U)} =  2W_{\rm low (u)}$ with
$W_{\rm low (u)}$ the low energy superpotential of $U(N/2+2)$
classically broken to $U(N/2 +1)\times U(1)$. The scales of the
theories are related by \refs{\CachazoJY,\CachazoZK}
\eqn\remul{ \Lambda_u^{N+4} = \eta \Lambda_U^{N+4}.}
with $\eta^2 = 1$. This is the source of the doubling of vacua
discussed at the end of section 2.2. However, note that \ffnn\ and
\mapu\ imply that $\Lambda^3$, which is the expansion parameter in
the $Sp(N)$ theory, is invariant under the $\IZ_2$ action $\eta\to
-\eta$.

\bigskip
\noindent{\it Effective superpotential for $S$}

Finally, in the case of the unbroken $Sp(N)$ vacuum, an effective
superpotential for $S$ can be easily computed by integrating it
in. This is done by performing a Legendre transform in two steps.
First, we introduce the intermediate superpotential
\eqn\legt{W_{\rm int}(S,\Lambda_H , C, g, m , \lambda ) = W_{\rm
low}(C, g , m , \lambda ) + (N+4) S \log{\Lambda_H \over C}.}
Second, we integrate out $C$ and use the matching relation \mapu\
to get
\eqn\effsu{W_{\rm eff}(S,\Lambda) = S\left[ \log \left(
\Lambda^{3(N/2+1)}\over S^{(N/2+1)} \right) + (N/2+1) \right] +
\sum_{k=2}^{\infty} a_k S^k.}

\newsec{Examples}

In this section we consider the six cases studied in \KrausJF.
Namely, $Sp(4)$, $Sp(6)$ and $Sp(8)$ with $\Phi$ in the
antisymmetric traceful representation and with $\Phi$ in the
antisymmetric traceless representation. The theory is deformed by
a cubic tree level superpotential and studied around the unbroken
classical vacuum.

As a consistency check of our approach we also include at the end
of this section the analysis of $Sp(0)$ and $Sp(2)$.

The tree level superpotential is given by \cusp,
\eqn\cusp{ W_{\rm tree} = {g\over 3}\Tr \Phi^3 + {m\over 2}\Tr
\Phi^2 + \lambda \Tr\Phi.}

In the traceful case we set $\lambda = 0$ and for the traceless
case we use it as a Lagrange multiplier imposing the vanishing of
$\langle \Tr \Phi \rangle$.

The strategy to compute the low energy superpotential for $Sp(4)$,
$Sp(6)$ and $Sp(8)$ is based on the integrability of \kqwp, which
was proven in the section 2.3. $W_{\rm low}$ is obtained by first
computing $\half \langle \Tr \Phi^2_u \rangle$ in the $U(N/2+2)$
theory classically broken to $U(N/2+1)\times U(1)$. Then \hjps\ is
used to get $\half \langle \Tr \Phi^2_U \rangle = \langle \Tr
\Phi^2_u \rangle$. The $Sp(N)$ vev $\half \langle \Tr \Phi^2
\rangle$ is obtained from the third equation in \cubr. Finally,
$\half \langle \Tr \Phi^2 \rangle$ is integrated with respect to
$m$ to get $W_{\rm low}$.

The explicit computation is carried out in appendix A. The results
come out in terms of $\Lambda_U$. Therefore we have to find the
exact relation between the scales of the theories. Using \ffnn\
and \mapu\ we find,
\eqn\comff{ \Lambda_U^{N+4} = g^{-2} m^{1-N/2}\Lambda^{3(N/2+1)}}
up to a numerical constant.

Comparing the leading order term of the superpotentials in
appendix A
\eqn\lead{ W_{\rm low} = (N+2){m^3\over g^2}\left( g\Lambda_U\over
m \right)^{2(N+4)/(N+2)} + {\cal O}(\Lambda^{4(N+4)/(N+2)}) }
with the standard definition of $\Lambda$, i.e.,
\eqn\stad{ W_{\rm low} = (N/2+1)\Lambda^3 + {\cal O}(\Lambda^6), }
the exact relation turns out to be
\eqn\matr{ \Lambda^{N+4}_{U} = g^{-2} m^{1-N/2}\left(
{\Lambda^3\over 2}\right)^{N/2+1}.}

The final result of using \matr\ in the superpotentials of
appendix A are listed below.

Note that $W_{\rm eff}(S)$ can easily be computed using \legt\ but
we will not do it here.

\bigskip
\noindent{\it $Sp(4)$ low energy superpotential:}

\noindent Traceful case:
\eqn\fspf{ W_{\rm low} = 3\Lambda^3 - \half \Lambda^6 {g^2\over
m^3}-\half \Lambda^9{g^4\over m^6} -{187\over
216}\Lambda^{12}{g^6\over m^9} -{1235\over
648}\Lambda^{15}{g^8\over m^{12}}+ {\cal O}(\Lambda^{18}). }
Traceless case:
\eqn\fuqw{ W_{\rm low} = 3 \Lambda^3 + {\cal O}(\Lambda^{18}). }

\bigskip
\noindent{\it $Sp(6)$ low energy superpotential:}

\noindent Traceful case:
\eqn\suplee{W_{\rm low} = 4\Lambda^3 -{3\over 2}\Lambda^6{g^2\over
m^3} -{47\over 24}\Lambda^9{g^4\over m^6} -{75\over
16}\Lambda^{12}{g^{6}\over m^9} -{7437\over
512}\Lambda^{15}{g^{8}\over m^{12}} + {\cal O}(\Lambda^{18}).}
Traceless case:
\eqn\fuqw{ W_{\rm low} = 4\Lambda^3 -{1\over 6}\Lambda^{6}{g^{2}
\over m^{3}} -{7\over 216}\Lambda^{9}{g^{4} \over m^{6}} -{5\over
432}\Lambda^{12}{g^{6} \over m^{9}} -{221\over
41472}\Lambda^{15}{g^{8} \over m^{12}} + {\cal O}(\Lambda^{18}). }

\bigskip
\noindent{\it $Sp(8)$ low energy superpotential:}

\noindent Traceful case:
\eqn\suplee{W_{\rm low} = 5\Lambda^3 - {5\over
2}\Lambda^6{g^2\over m^3} -{13\over 4}\Lambda^9{g^4\over m^6} -
{65\over 8}\Lambda^{12}{g^{6}\over m^9} - {2147\over
80}\Lambda^{15}{g^{8}\over m^{12}} + {\cal O}(\Lambda^{18}). }
Traceless case:
\eqn\fuqw{ W_{\rm low} = 5\Lambda^3 - {1\over 4}\Lambda^6{g^2\over
m^3}-{1\over 10}\Lambda^9{g^4\over m^6} -{7\over
100}\Lambda^{12}{g^{6}\over m^9} - {1\over
16}\Lambda^{15}{g^{8}\over m^{12}} + {\cal O}(\Lambda^{18}). }

Comparing these results with the ones obtained by using the
dynamically generated superpotentials reviewed in appendix B, we
find complete agreement to all orders computed.

\bigskip
\noindent{\it $Sp(0)$ and $Sp(2)$ low energy superpotentials. A
consistency check:}

$Sp(0)$ and $Sp(2)$ are special. The former should clearly give a
trivial result. The latter is interesting since in our convention
$Sp(2)\simeq SU(2)$ and the antisymmetric matter is actually a
singlet of the gauge group. This means that it does not
participate in the strong coupling dynamics of the theory. This is
why $Sp(2)$ is a non trivial consistency check of our formalism.

According to the general analysis of section 2, we are instructed
to consider $U(4)\to U(2)\times U(2)$ and $U(6)\to U(4)\times
U(2)$ respectively. Compute $W_{\rm low(U)}$ in the vacua with
$t=2$ and use it in \ansz. Since $t=2$ we only have to consider
$U(2)\to U(1)\times U(1)$ and $U(3)\to U(2)\times U(1)$
respectively.

In the case of $U(2)\to U(1)\times U(1)$ it is known that $S_u=0$
\CachazoPR\ and therefore the low energy superpotential does not
depend on $\Lambda_u$. This implies that $W_{\rm low(u)}$ can be
evaluated at any $\Lambda_u$ in particular, at $\Lambda_u = 0$ to
get the classical answer,
\eqn\arre{W_{\rm low(u)}(\Lambda_u, g_u, m_u, \lambda_u) = {1\over
6}{m^3_u\over g_u^2} - {m_u\lambda_u\over g_u}}
where $g_u=h_2$, $m_u=h_1$ and $\lambda_u =h_0$ in \duasu.

Using this in \ansz\ we find\foot{Recall that $W_{\rm low(U)} =
2W_{\rm low(u)}$. }
\eqn\guev{ W_{\rm low} = 0.}
This is indeed the correct result.

In the case $U(3)\to U(2)\times U(1)$, the low energy
superpotential is given by \refs{\ElitzurGK,\CachazoJY},
\eqn\sptw{ W_{\rm low(u)}(\Lambda_u, g_u , m_u , \lambda_u=0) =
{1\over 6}{m_u^3\over g_u^2} + 2 g_u \Lambda_u^3.}
Using this in \ansz\ together with the matching relation \matr\ we
get,
\eqn\losu{W_{\rm low} = 2\Lambda^3.}
This is the correct answer for pure $\N=1$ $SU(2)$ gauge theory as
can be seen from \loon.

\newsec{Conclusions and open questions}

The low energy superpotential of $\N=1$ $Sp(N)$ theories with
matter in the antisymmetric tensor representation and tree level
superpotential was computed using the generalized Konishi anomaly
equations found in \KrausJV\ and \AldayGB. A key role was played
by the generating function of chiral operators $T(z)$. Quantum
mechanically, $T(z)dz$ becomes a meromorphic differential on a
Riemann surface.

A remarkably simple condition on $T(z)dz$ was imposed which
accounts for the full IR dynamics. The condition is the
integrality of its $A$- and $B$-periods. Based on all the examples
in the literature and the ones presented here, it is reasonable to
propose that this is always the case in theories where a
hyperelliptic Riemann surface emerges. One possible explanation is
the following: In such theories a string theory realization with a
dual involving fluxes might be available. On the dual side, the
fluxes through compact cycles are quantized and are given as the
periods of a closed real form. In the projection to a Riemann
surface\foot{The compact cycles are in general embedded in
manifolds of complex dimension three. If two of them are trivial,
then a projection to a Riemann surface is possible.} this form
gives rise to a one form with integer periods. However, the one
form is real instead of meromorphic. It is only on-shell that this
real one form becomes meromorphic and agrees with $T(z)dz$. On the
other hand, $T(z)dz$ is meromorphic by definition but the
integrality of its periods is valid only on-shell. It would be
interesting to explore the full range of validity of this dual
description.

In the process of imposing the integrality condition of the
periods of $T(z)dz$ in the $Sp(N)$ theory, we found a very precise
map to a $U(N+2n)$ theory. The map was given for a general degree
$n+1$ tree level superpotential. In the vacuum where $Sp(N)$ is
classically broken to $\prod_{i=1}^n Sp(N_i)$, $U(N+2n)$ is broken
to $\prod_{i=1}^nU(N_i+2)$.

However, one point we did not discuss is the following: In the IR,
$U(N+2n)$ has low energy group $U(1)^n$. Their couplings are
frozen and depend on the parameters in the superpotential and
$\Lambda_U$. What is the interpretation of this on the $Sp(N)$
side?

In section 2 we found a very simple expression \ansz\ for the low
energy superpotential $W_{\rm low}$ of $Sp(N)$ with a cubic tree
level superpotential. It would be interesting to generalize it to
tree level superpotentials of arbitrary degree.

Generalizations of the approach presented in this work to $Sp(N)$
and $SO(N)$ with symmetric/antisymmetric tensors and fundamentals
is surely possible and interesting. We leave it for future work.

As mentioned in the introduction we have shown that the vanishing
of the period of $T(z)dz$ through a given cut {\it does not} imply
that the cut closes up on-shell. Also surprising is the opposite
case, namely, the period of $T(z)dz$ is non zero but the cut
closes up on-shell. This does not arise in the cases studied in
this work but we believe it will show up for $SO(N)$ with a
symmetric tensor. This is currently under investigation.

We also would like to comment on the role of the term ${d\over
dz}R(z)$ in the anomaly equation for $T(z)$. Its presence prevents
a given cut from closing up on-shell when the period of $T(z)dz$
is zero through it as one would naively have expected. It would be
interesting to get a more geometrical explanation of this
phenomenon. In terms of the $U(N/2+n)$ theory, the role of such a
term is to produce a classical breaking pattern of the form
$\prod_{i=1}^nU(h_i)$ where $h_i$ is the dual Coxeter number of
the original $Sp(N_i)$ factor supported at the i$^{th}$ cut. It is
reasonable to think that this is generic and it would be
interesting to explore it in other examples which might include
exceptional groups.

Finally, we believe that the techniques used here are a reliable
and simple way to study the IR dynamics of supersymmetric field
theories and should be explored in more detail.

\bigskip
\centerline{\bf Acknowledgements}

We are particularly indebted to E. Witten for very useful
discussions. It is also a pleasure to thank B. Feng, R. Gopakumar,
O. Lunin, N. Seiberg, and D. Shih for helpful conversations. This
work was supported by DOE grant \#DE-FG02-90ER40542 to IAS.

\appendix{A}{$U(N+4)$ results}

In this appendix we review and derive the relevant $U(N+4)$ field
theory results needed to compute the low energy superpotential
$W_{\rm low}$ of the $Sp(N)$ theory discussed in section 3. The
theory has a cubic tree level superpotential
\eqn\lwexx{ W_{\rm tree(U)} = {g_u\over 3}\Tr \Phi_U^3 + {m_u\over
2}\Tr \Phi_U^2 + \lambda_u \Tr\Phi_U,}
and it is classically broken to $U(N+2)\times U(2)$. Here we use
the notation $g_u$, $m_u$ and $\lambda_u$ for both the $U(N+4)$
and the $U(N/2+2)$ theories as they share the same tree level
superpotential.

There are two possible ways to proceed. We will briefly discuss
the first one, which is very powerful and can be applied to
arbitrary high values of $N$. However, this is not what we use
since the three cases considered in section 3 map to three cases
already considered in \CachazoZK\ using a strong coupling
analysis. The reader can skip the first one unless interested in
larger values of $N$ for which the strong coupling analysis is
complicated.

\bigskip
\noindent{\it Matrix model effective superpotential}

The first method is to use the matrix model
\refs{\DijkgraafFC,\DijkgraafVW,\DijkgraafDH} to compute an
effective superpotential for $U(N+4)$ around the classical vacuum
$U(N+2)\times U(2)$. This superpotential is a function of $S_1 =
-{1\over 4\pi i}\oint_{A_1} y_U(z)dz$ and $S_2 = -{1\over 4\pi
i}\oint_{A_2} y_U(z)dz$ given by
\eqn\effmm{ \eqalign{ W_{\rm eff}(S_1,S_2) =\ &
-{\half}(N+2)\int_{\hat B_1^r}y_U(z)dz - \int_{\hat
B_2^r}y_U(z)dz+ (N+4) W(\Lambda_0) \cr & - 2 (N+4) S \log \left(
-{\Lambda_0\over \Lambda_U } \right) + 2\pi i (b_1 S_1+ b_2 S_2) ,
}}
where $\hat B_i^r$ are regularized non-compact cycles, $\Lambda_0$
is a cut off, which should be taken to infinity at the end of the
computation and $b_i$'s are integers (for more details see e.g.
\CachazoYC).

At the extremum of \effmm\ we get $W_{\rm low(U)} = W_{\rm
eff}(<S_1>, <S_2>)$. This result can be used in \ansz\ to get the
$Sp(N)$ low energy superpotential $W_{\rm low}$.

\bigskip
\noindent{\it Strong coupling approach}

The second method is to use a strong coupling analysis where the
tree level superpotential is thought of as a deformation of a
$\N=2$ $U(N+4)$ theory. Since we are only interested in vacua with
confinement index $t=2$ we can consider a $\N=2$ $U(N/2+2)$
theory. The only $\N=1$ supersymmetric vacua are located at points
in the $\N=2$ Coulomb moduli space that satisfy
\eqn\conp{ P^2_{{\tilde N}}(z) - 4\Lambda_u^{2\tilde N} = {1\over
g_u^2}(W_u'^2(z)+ f_u(z)) H^2_{{\tilde N}-n}(z),}
where $\tilde N = N/2+2$ and
\eqn\ppp{ P_{\tilde N}(z) = \det (z\id -\Phi_u ) = z^{\tilde N} -
\langle\Tr \Phi_u\rangle z^{{\tilde N}-1} +\left( \half \langle\Tr
\Phi_u\rangle^2 -\half \langle\Tr \Phi_u^2\rangle \right)
z^{{\tilde N}-2} + ... }

As shown in section 2, the cases $Sp(4)$, $Sp(6)$ and $Sp(8)$ map
to $U(8)\to U(6)\times U(2)$, $U(10)\to U(8)\times U(2)$, and
$U(12)\to U(10)\times U(2)$, respectively. Fortunately, as
mentioned above we are interested in the vacua with confinement
index $t=2$. Therefore, all we need to consider is $U(4)\to
U(3)\times U(1)$, $U(5)\to U(4)\times U(1)$, and $U(6)\to
U(5)\times U(1)$, respectively. The problem of finding $P_{\tilde
N}(z)$, $f_u(z)$ and $H_{{\tilde N} - n}(z)$ satisfying \conp\ was
solved for these three cases in \CachazoZK, from where we borrow
the results\foot{$U(5)\to U(4)\times U(1)$ was first studied in
\CachazoJY. $U(4)\to U(3)\times U(1)$ was also studied in
\FerrariYR.}.

In this appendix we introduce a subscript ${\rm Sp}$ to all
$Sp(N)$ quantities in order to avoid possible confusions.

\bigskip
\noindent{\it $U(4)$ case:}

The solution to \conp\ is:
\eqn\ufdt{P_4(z) = (z-a)^2((z+a)^2+ v (z+2 a))- 2\Lambda_u^4 \quad
{\rm with } \quad a^3 = {\Lambda_u^4\over v}.}
Clearly, in the semiclassical limit $\Lambda_u\to 0$, $a\to 0$ and
$P_4(z) \to z^3(z+v)$, showing that here $U(4)$ is broken to $U(3)
\times U(1)$.

{}From \conp\ and \ufdt, we find that
\eqn\ufowpc{ {1\over g_u}W'(z)= z^2 + v z-a^2.}

In this computation the freedom to shift $z$ was used. In order to
recover this degree of freedom we shift $z\to z+\delta$. Note that
$W'(z)$ in \ufowpc\ depends on $\delta$, $v$ and $a$. Comparing it
to $W'(z) = g_u z^2 + m_u z + \lambda_u$ we get two equations
which together with the constraint $a^3 = {\Lambda_u^4\over v}$,
determine $\delta$, $v$ and $a$ as functions of $m_u/g_u$,
$\lambda_u/g_u$ and $\Lambda_u$.

These equations can be solved in a power expansion in $\Lambda_u$
around $\Lambda_u=0$. We also choose to expand around the
classical solution
\eqn\clasx{ v_{\rm cl}=\sqrt{\left({m_u\over
g_u}\right)^2-4{\lambda_u\over g_u}} .}

Consider first the traceful case. We set $\lambda_u =0$ and use
the values of $\delta$, $a$ and $v$ to compute $P_4(z)$. It is
important to remember the shift $z\to z+\delta$. From $P_4(z)$ and
\ppp\ it is easy to compute $\langle \Tr \Phi_u^2 \rangle$ which
is equal to $\half\langle \Tr \Phi_U^2 \rangle$. Using this last
result in \cubr\ we get
\eqn\ttr{ \half\langle \Tr \Phi^2 \rangle_{\rm Sp} = {m^2_u\over
g^2_u}\left( 2 T+{14\over 3}T^2+20 T^3+{8602\over
81}T^4+{153140\over 243}T^5 + {\cal O}(T^6) \right) }
with
\eqn\tspf{T = \left( {g_u\Lambda_u\over m_u}\right) ^{8/3}.}

Finally, note that only the ratio $m_u/g_u$ appears and can be
replaced by $m/g$. Recall that the $Sp(N)$ tree level
superpotential is given by \cusp\ and it is proportional to
\lwexx. $W_{\rm low}$ can then be obtained by integrating \ttr\
with respect to $m$,
\eqn\sinqw{W_{\rm low(Sp)}^{\rm Traceful} = {m^3\over g^2}\left( 6
T - 2T^2 -4T^3 - {374\over 27}T^4 - {4940\over 81}T^5 +{\cal
O}(T^6) \right) }
with
\eqn\tspx{T = \left( {g\Lambda_u\over m}\right) ^{8/3}.}

For the traceless case we have to use $\lambda$ to set the second
equation in \cubr\ to zero, i.e.,
\eqn\trss{ \langle \Tr \Phi_U \rangle + 2{m_u\over g_u} = 0  }
or equivalently,
\eqn\trsst{\langle \Tr \Phi_u \rangle + {m_u\over g_u} = 0.}
The solution to \trsst\ is
\eqn\solll{ {\lambda_u\over g_u} = -{m^2_u\over g^2_u} T + {\cal
O}(T^6). }
Using this and following the same procedure as before we compute
\eqn\qwgh{ \half\langle \Tr \Phi^2 \rangle_{\rm Sp} ={m^2_u\over
g_u^2}\left( T + {\cal O}(T^6) \right) .}
Integrating with respect to $m$ we find
\eqn\sjqw{ W_{\rm low(Sp)}^{\rm Traceless} = {m^3\over g^2}\left(
6 T + {\cal O}(T^6)\right) }
with $T$ given by \tspx.

\bigskip
\noindent{\it $U(5)$ case:}

The solution to \conp\ is:
\eqn\cocp{P_5(z)= (z^2+a z - 2 a c)^2(z+c) - 2\Lambda_u^5 \quad
{\rm with } \quad a^2 = {\Lambda_u^5\over c^3.} }
In the semiclassical limit $\Lambda_u \to 0$, $a\to 0$ and
$P_5(z)\to z^4 (z+c)$, showing that here $U(5)$ is broken to
$U(4)\times U(1)$.

{}From \conp\ and \cocp, we find that
\eqn\soll{ {1\over g_u}W'(z) =  z^2+(a+c)z-ac.}

Following the same steps as in the $U(4)$ case, we expand around
the classical solution
\eqn\clasy{ c_{\rm cl}=\sqrt{\left({m_u\over
g_u}\right)^2-4{\lambda_u\over g_u}}.}

In the traceful case with $\lambda_u=0$ we get
\eqn\psxw{\half\langle \Tr \Phi^2 \rangle_{Sp} = {m_u^2\over
g^2_u}\left( 4 T + 12 T^2 + {141\over 2} T^3 + {525}T^4 +
{141303\over 32}T^5 +  {\cal O}(T^6) \right) }
with $T=(g_u \Lambda_u / m_u)^{5/2}$.

Integrating with respect to $m$ after replacing $m_u/g_u$ by $m/g$
we get,
\eqn\aswwq{ W_{\rm low(Sp)}^{\rm Traceful} = {m^3\over g^2}\left(
8 T - 6 T^2 -{47\over 3}T^3 - 75T^4 - {7437\over 16}T^5 + {\cal
O}(T^6) \right) }
with $T=(g \Lambda_u / m)^{5/2}$.

For the traceless case we have to solve \trsst\ to get,
\eqn\trlam{ {\lambda_u\over g_u} = {m^2_u\over g_u}\left( -{4\over
3}T -{4\over 9}T^2 -{7\over 18}T^{3} -{35\over 81}T^4- {4199\over
7776}T^5  + {\cal O}(T^6)\right) .}

Using this we compute
\eqn\hcar{\half\langle \Tr \Phi^2 \rangle_{Sp} = {m_u^2\over
g_u^2}\left( 4 T  + {4\over 3}T^2 + {7\over 6}T^3 + {35\over
27}T^4 + {4199\over 2592}T^5 + {\cal O}(T^6) \right) . }

Integrating with respect to $m$ we get,
\eqn\aspq{ W_{\rm low(Sp)}^{\rm Traceless} = {m^3\over g^2}\left(
8 T - {2\over 3}T^2 - {7\over 27}T^3 - {5\over 27}T^4 -{221\over
1296}T^5 + {\cal O}(T^6) \right) . }

\bigskip
\noindent{\it $U(6)$ case:}

The solution to \conp\ is:
\eqn\Coubranch{ \eqalign{ P_6(z) =\ & \left[z^2 +
(h+g)z+{(3h+g)(9h^3+15h^2g -hg^2+g^3)
 \over 108 h^2}\right]^2 \cr & \left[z^2 -{(h-g)(3h-g)^2(3h+g)\over 108
 h^2}\right] -2\Lambda_u^6}}
with $g$ and $h$ satisfying the constraint
 \eqn\constraint{g^5(g^2-9h^2)^2 = 27^3h^3\Lambda_u^6.}

The classical limit $\Lambda_u\to 0$ with $g\to 0$ gives $P_6(z)
\to (z+{h\over 2})^5 (z-{h\over 2})$; i.e.\ $U(6) \to U(5) \times
U(1)$.

\eqn\wcouf{
 {1\over g_u}W'(z) = z^2 + {2g\over 3} z + {g^4 - 6g^2 h^2 - 27 h^4 \over
 108 h^2}.}

Following again the same steps as in the $U(4)$ case, we expand
around the classical solution
\eqn\classw{ h_{\rm cl} =-\sqrt{\left( { m_u\over
g_u}\right)^2-4{\lambda_u\over g_u} }.}

In the traceful case with $\lambda_u=0$ we get
\eqn\psxq{\half\langle \Tr \Phi^2 \rangle_{Sp} = {m^2_u\over
g^2_u}\left( 6 T + 18 T^2 +{546\over 5}T^3 + 858 T^4 + {38646\over
5} T^{5} + {\cal O}(T^{6}) \right) }
with $T=(g_u\Lambda_u/m_u)^{12/5}$.

Integrating with respect to $m$,
\eqn\gwfi{W_{\rm low(Sp)}^{\rm Traceful} = {m^3\over g^2}\left( 10
T -10 T^2 -26 T^3 -130 T^4 - {4294\over 5} T^{5} + {\cal O}(T^6)
\right) .}

For the traceless case we have to solve \trsst\ to get,
\eqn\pklam{ {\lambda_u\over g_u} = {m^2_u\over g^2_u}\left(
-{3\over 2}T - {9\over 20}T^2 - {21\over 25}T^3 - {231\over
125}T^4 - {9\over 2}T^5 + {\cal O}(T^6) \right) .}

Using this we compute
\eqn\hcar{\half\langle \Tr \Phi^2 \rangle_{Sp} = {m^2_u\over
g_u^2}\left( 6 T + {9\over 5} T^2 + {84\over 25} T^3 + {924\over
125} T^4 + {18}T^5 + {\cal O}(T^6) \right) .}

Integrating with respect to $m$,
\eqn\popk{W_{\rm low(Sp)}^{\rm Traceless} = {m^3\over g^2}\left(
10 T - T^2 - {4\over 5}T^3 -{28\over 25}T^4 -{2}T^5 + {\cal
O}(T^6) \right) .}

\appendix{B}{$Sp(N)$ results from $W_{\rm dyn}$}

Exact results for the dynamically generated superpotential of
$\N=1$ $Sp(4)$ and $Sp(6)$ gauge theories with one traceless
antisymmetric tensor and some number of flavors were computed in
\ChoBI\ and \CsakiEU. Very recently, results for $Sp(8)$ and for a
traceful antisymmetric tensor were obtained in \KrausJF\ by
applying the general strategy of \CsakiEU.

Here we will only need the results for the case without flavors.
In \KrausJF, a cubic tree level superpotential,
\eqn\krau{ W_{\rm tree} = {g\over 3}{\tilde O}_3 + {m\over
2}{\tilde O}_2 }
where ${\tilde O}_i =\Tr \Phi^i$, was added to $W_{\rm dyn}$ and
the F-term equations were solved. Those are the results we will
write here.

{\it $Sp(4):$}

The dynamically generated superpotential is
\eqn\spft{ W_{\rm dyn} = {2\sqrt{2}\over
\sqrt{m}}{\Lambda^{9/2}\over  {\tilde O}^{1/2}_2 } . }

The solution to the F-term equations leads to
\eqn\sopf{ \eqalign{ W_{\rm low}^{\rm Traceless} &= 3\Lambda^3,
\cr W_{\rm low}^{\rm Traceful} &= 3\Lambda^3 - {1\over
2}\Lambda^6{g^2\over m^3} -{1\over 2}\Lambda^9{g^4\over m^6}
-{187\over 216}\Lambda^{12}{g^6\over m^9} - {1235\over
648}\Lambda^{15}{g^8\over m^{12}} + {\cal O}(\Lambda^{18}).} }
Here we fix a misprint in \KrausJF\ in the coefficient of the term
$\Lambda^{15}$ of $W_{\rm low}^{\rm Traceful}$.

{\it $Sp(6):$}

The dynamically generated superpotential is
\eqn\spss{W_{\rm dyn} = {8\Lambda^6 \over m {\tilde O}_2 \left( (
\sqrt{R} + \sqrt{R+1})^{2/3} + (\sqrt{R} + \sqrt{R+1})^{-2/3} - 1
\right)} }
where $R = -12 {\tilde O}_3^2/ {\tilde O}_2^3$.

The solution to the F-term equations leads to
\eqn\sopqw{ \eqalign{ W_{\rm low}^{\rm Traceless} &= 4\Lambda^3-
{1\over 6}\Lambda^6{g^2\over m^3} -{7\over 216}\Lambda^9{g^4\over
m^6} -{5\over 432}\Lambda^{12}{g^{6}\over m^9} - {221\over
41472}\Lambda^{15}{g^{8}\over m^{12}}  + {\cal O}(\Lambda^{18}),
\cr W_{\rm low}^{\rm Traceful} &= 4\Lambda^3 - {3\over
2}\Lambda^6{g^2\over m^3} -{47\over 24}\Lambda^9{g^4\over m^6}
-{75\over 16}\Lambda^{12}{g^{6}\over m^9} - {7437\over
512}\Lambda^{15}{g^{8}\over m^{12}} + {\cal O}(\Lambda^{18}).} }

{\it $Sp(8):$}

The dynamically generated superpotential is
\eqn\spseight{ W_{\rm dyn} = K \left(  -36 R_4 + 144 b^2 R_4 + 288
c R_4 + 8 R_3^2 + 192 b c R_3 + 1152 b^2 c^2 - 36 b^2 - 72 c+9
\right)^{-1}, }
where $K =24\Lambda^{15/2}/  (m{\tilde O}_2)^{3/2}$, $R_3 =
{\tilde O}_3/{\tilde O}_2^{3/2}$ , $R_4 = {\tilde O}_4/{\tilde
O}_2^2$, and with $b$ and $c$ solutions of the following set of
polynomial equations
\eqn\slpoy{ \eqalign{ 12 R_4 + 16 b R_3 - 192 b^2 c+ 24 b^2 + 96
c^2 -3 &=0, \cr 12 b R_4+ 8 b^2 R_3 + 8 R_3 c - 96 b c^2 + 24 b c-
3 b &=0. } }

The solution to the F-term equations leads to
\eqn\sopqw{ \eqalign{ W_{\rm low}^{\rm Traceless} &= 5\Lambda^3-
{1\over 4}\Lambda^6{g^2\over m^3} -{1\over 10}\Lambda^9{g^4\over
m^6} -{14\over 200}\Lambda^{12}{g^{6}\over m^9} - {1\over
16}\Lambda^{15}{g^{8}\over m^{12}} + {\cal O}(\Lambda^{18}), \cr
W_{\rm low}^{\rm Traceful} &= 4\Lambda^3 - {5\over
2}\Lambda^6{g^2\over m^3} -{13\over 4}\Lambda^9{g^4\over m^6}
-{65\over 8}\Lambda^{12}{g^{6}\over m^9} - {2147\over
80}\Lambda^{15}{g^{8}\over m^{12}}  + {\cal O}(\Lambda^{18}).} }

\listrefs
\end